\documentclass{aa501}
\usepackage{graphicx}

\begin{document}
\title{Density profiles in a spherical infall model with non--radial motions}

   \subtitle{}

   \author{N. Hiotelis
\inst{ }
   \fnmsep\thanks{\emph{Present address:} Roikou 17-19,
               Neos Kosmos, Athens, 11743 Greece}
}
 \offprints{N. Hiotelis}

   \institute{Lyceum of Naxos, Chora of Naxos, Naxos 84300, Greece\\
              email: hiotelis@avra.ipta.demokritos.gr}

    \date{Received  3 September 2001 / Accepted 5 November 2001}
   \abstract{A generalized version of the Spherical Infall Model
    (SIM) is used to study the effect of angular momentum
    on the final density profile of a spherical structure.
   The numerical method presented is able to handle a variety
   of initial density profiles (scale or not scale free) and no assumption
     of self-similar evolution is required.
    The realistic initial overdensity profiles used are derived by a CDM power spectrum.
We show that the amount of angular momentum  and the initial
overdensity profile affect the slope of the final density profile
at the inner regions.
 Thus, a larger amount of angular momentum or shallower initial
  overdensity profiles lead to shallower final density profiles at the
  inner regions.
On the other hand, the slope at the outer regions
   is not affected by the amount of
   angular momentum and has an almost constant value equal to that
   predicted in the radial collapse case.
   }

   \maketitle
%

\section{Introduction}

It is likely that dark matter halos are formed by the evolution of
small density perturbations in the early Universe. The matter
contained in a perturbed region progressively detaches from the
general flow and  after reaching a radius of maximum expansion it
collapses to form an individual  structure. The most simple case
is when this region is spherical, isolated and undergoing a radial
collapse. This is the spherical infall model (hereafter SIM). SIM
has been extensively discussed in the literature (Gunn \& Gott
\cite{gunn72}; Gott \cite{gott}; Gunn \cite{gunn77}; Fillmore \&
Goldreich \cite{fillmore}; Bertschinger \cite{bert}; Hoffman \&
Shaham \cite{hoffman}, hereafter HS; White \& Zaritsky
\cite{white}).

 The final density profile after a collisionless
evolution of the matter depends on its initial density profile and
the underlying cosmology. Self-similarity solutions (Fillmore \&
Goldreich \cite{fillmore}; Bertschinger \cite{bert}, HS) show that
a power-law initial density profile relaxes to a final density
profile given by $\rho(r)\propto r^{-\alpha}$ with $\alpha\geq 2$.
Furthermore, recent numerical studies that relax the assumption of
self-similarity,  also give final density profiles steeper than
$r^{-2}$. Lokas \& Hoffman (\cite{lokas01}) found values of
$\alpha$ in the range 2 to 2.3.

The density profiles of galactic halos do not seem to follow power
laws. Numerical studies (Quinn et al. \cite{quinn}; Frenk et al.
\cite{frenk}; Dubinski \& Galberg \cite{dubinski}; Crone et al.
\cite{crone}; Navarro et al. \cite{navarro}; Cole \& Lacey
\cite{cole}; Huss et al. \cite{huss}; Fukushinge \& Makino
\cite{fuku}; Moore et al. \cite{moore}; Jing \& Suto \cite{jing})
showed that the profile of relaxed halos steepens monotonically
with radius. The logarithmic slope
$\alpha=-\frac{\mathrm{d}\ln\rho}{\mathrm{d}\ln r}$ is less than 2
near the center and larger than 2 near the virial radius of the
system. The value of $\alpha $ near the center of the halo is not
yet known. Navarro et al. (\cite{navarro}) claimed $\alpha=1$
while Kravtsov et al. (\cite{kravtsov}) initially claimed
$\alpha\sim 0.7$ but in their revised conclusions (Klypin et al.
\cite{klypin}) they argue that the inner slope varies from 1 to
1.5. Moore et al. (\cite{moore}) found a slope $\alpha=1.5$ at the
inner regions of their N-body systems.\\
\indent In this paper we study the final density profiles
predicted by the SIM, when non-radial motions are included.
Studies concerning the role of non-radial motions  have been
presented by Ryden \& Gunn (\cite{ryden87}), Ryden
(\cite{ryden88}), Gurevich \& Zybin (\cite{gura}, \cite{gurb}),
Avila-Reese et al. (\cite{avila}), White \& Zaritsky
(\cite{white}), Sikivie et al.(\cite{siki}) and recently by Nusser
(\cite{nusser}).\\
\indent In Sect. 2 we discuss the SIM and the associated problems.
In Sect. 3 the description of the numerical method as well as the
way the angular momentum is included  is given. The initial
conditions and the results are given in Sect. 4 and are summarized
in Sect. 5.

\section{The spherical infall model (SIM)}

SIM is based on the physical process described in Gunn
(\cite{gunn77}) and in Zaroubi \& Hoffman (\cite{zaroubi}): In an
expanding spherical region the maximum expansion radius $\zeta$
(apapsis) of a shell is a monotonic increasing function of its
initial radius $x_i$ and is given by the relation:

\setcounter{equation}{0}
\begin{equation}
\zeta=g(x_i)=\frac{1+\Delta_i(x_i)}{1-\Omega^{-1}
_i+\Delta_i(x_i)}x_i,\label{eqb1}\\
\end{equation}

\noindent where  $\Omega_i$ is the initial value of the density
parameter of the Universe and
 $\Delta_i$ is the relative excess of mass inside
the sphere of radius $x_i$, given by \

\begin{equation}
\Delta_i(x_i)=\frac{M(x_i)-M_b(x_i)}{M_b(x_i)}
=\frac{3}{x^3_{i}}\int_0^{x_i}x^2\delta_i(x)\mathrm{d}x.\label{eqb2}\\
\end{equation}

 In (\ref{eqb2}), $M$ is the mass of the spherical region,
 $M_ b$ is the mass of the unperturbed Universe
and $\delta_i$ is the spherically symmetric perturbation of the
density field ($\delta_i(x)=\frac{\rho(x)-\rho_{b,i}}{\rho_{b,i}}$
where $\rho$ is the density and $\rho_{b,i}$ is the constant
density of the homogeneous Universe at the initial conditions).
The time spent for a shell to reach its above turnaround radius
is:
\begin{equation}
t_{ta}=\frac{1+\Delta_i(x_i)}{2H_i\Omega^{1/2}_i
[1-\Omega^{-1}_i+\Delta_i(x_i)]^{\frac{3}{2}}}\pi,\label{eqb3}\\
\end{equation}
where $H_i$ is the value of Hubble's constant at the epoch of the
initial conditions. The above equations are valid for bound
shells, so the condition $1-\Omega^{-1}_i+\Delta_i(x_i)>0$ is
satisfied. A shell, after reaching its turnaround radius,
collapses, re-expands to a new (smaller) turnaround radius and so
on. The limiting value of this radius, after a large number of
such oscillations, is the final radius of the shell, corresponding
to the relaxed state of the system.
 In any reasonable potential  a shell spends most of its orbital time near the maximum radius
(Gunn \cite{gunn77}). So, if shell crossing -occurring during the
collapse stage- had no dynamical consequences, the final
distribution of mass could be approximated by the distribution
resulting if every shell stopped at its maximum expansion radius.
This should lead to a ``turnaround density profile'' $\rho_{ta}$
of the form

\begin{equation}
\rho_{ta}(\zeta)=\left(\frac{x_i}{\zeta}\right)^2\rho_i(x_i)
\left(\frac{\mathrm{d}\zeta}{\mathrm{d}x_i}\right)^{-1}.\label{eqb4}\\
\end{equation}

 \noindent In deriving (\ref{eqb4}) the conservation of mass is used
 ($M(x_i)=M(\zeta)$). This is an important relation
  since this distribution of mass is used as the initial one in SIM.
  SIM assumes that the collapse is gentle enough. This means that
   the orbital period of the inner
 shell is much smaller than the collapse time of the outer shells
  (Zaroubi \& Hoffman \cite{zaroubi}).
 This implies
 that the radial action $\oint v(r)\mathrm{d}r$,
  where $v$ is the radial velocity, is an adiabatic
 invariant of the inner shell.
 As the outer shells collapse, the potential changes slowly and
 because of the above adiabatic invariant, the inner shell
 shrinks. The collapse factor depends on the time the
 mass of the outer shells (passing momentarily) spends inside the inner shell.
Consider a shell with apapsis $\zeta$ and initial radius $x_i$.
The mass inside radius $\zeta$ is a sum of two components. The
first one, (permanent component, $M_p$), is due to the shells with
apapsis smaller than $\zeta$ and the second (additional mass,
$M_{add}$) is the contribution of the outer shells passing
momentarily through the shell $\zeta$. Because of the mass
conservation, the permanent component is given by the following
relation
\begin{equation}
M_p(\zeta)=M(x_i)=\frac{4}{3}\pi\rho_{b,i}x^3_{i}[1+\Delta_i(x_i)].\label{eqb5}\\
\end{equation}
 The additional component is:
\begin{equation}
M_{add}(\zeta)=\int_{\zeta}^R P_{\zeta}(x)
\frac{\mathrm{d}M(x)}{\mathrm{d}x}\mathrm{d}x.\label{eqb6}\\
\end{equation}
\noindent In (\ref{eqb6}), $R$ stands for the radius of the system
(the apapsis of the outer shell) and the distribution of mass
$M(x)$ is given by (\ref{eqb4}). $P_{\zeta}(x)$ is the probability
of finding the shell with apapsis $x$ inside radius $\zeta$,
calculated as the ratio of the time the outer shell (with apapsis
$x$) spends inside radius $\zeta$ to its period. In the general
case of non-radial collapse this ratio is given by the relation
$P_{\zeta}(x)=I(\zeta)/I(x)$ where

\begin{equation}
I(r)=\int_{x_p}^{r}\frac{\mathrm{d}n}{v_x(n)},
\end{equation}

\noindent where $x_p$ is the pericenter of the shell with apapsis
$x$ and $v_{x}(n)$ is the radial velocity of the shell with
apapsis $x$ as it passes from radius $n$. If the collapse is
radial then $x_p=0$. After the calculation of $M_{add}$ the
collapse factor $f(x_i)$ of a shell with initial radius $x_i$ and
apapsis $\zeta=g(x_i)$ is given by
\begin{equation}
f(x_i)=\frac{M_p(\zeta)}{M_p(\zeta)+M_{add}(\zeta)}.
\label{eqb8}
\end{equation}
 The final radius of
the shell is $x=f(x_i)\zeta$ and mass conservation leads to the
following final density profile

\begin{equation}
\rho(x)=\rho_{\mathrm{ta}}(\zeta)  f^{-3}(x_i) \left[1+\frac{\mathrm{d}\ln
f(x_i)}{\mathrm{d}\ln g(x_i)}\right]^{-1}. \label{eqb9}
\end{equation}
The potential energy of the system $W_{f}$ in the relaxed state is
related to its total energy by the virial theorem $W_f=2E$. The
energy of the system in a radial collapse is that at the
turnaround epoch when all shells are assumed to have zero
velocities at the same time. Thus $W_f=2W_{ta}$. A simple collapse
factor satisfying this requirement is $f=0.5$ for every shell,
 leading to similarity solutions (a final
density profile parallel in log-log space to that of the
turnaround epoch). However, the collapse factor is not constant.
N-body simulations (e.g. Voglis et al. \cite{voglis}, hereafter
VHH) show that $f$ is an increasing function of the initial radius
(or the turnaround radius) of the shell and its form is related to
the initial profile of the density perturbation.

In a radial collapse case eqn.(\ref{eqb8}) gives $f\rightarrow 0$
as $x_i\rightarrow 0$ resulting in very condensed central regions
with very steep density profiles (Lokas \cite{lokas00}). As it is
shown by Lokas \& Hoffman (\cite{lokas01}) the inner slope of the
density profile is
 between $2$ and $2.3$ even for low initial density peaks, far from
the values derived from N-body simulations that range from $1$ to $1.5$ for the
inner regions. On the other hand, the collapse factor in N-body simulations is
hardly less than 0.05, even in the very central regions (VHH). This difference is
easy to understand because no radial collapse exists in N-body simulations. During
the expansion and the early stage of collapse particles acquire random velocities
that prevent them from penetrating the inner regions. The consequence is the
reduction of $M_{add}$ in (\ref{eqb8}) that leads to larger values of $f$. In this
sense the approximations based on constant $f$ may be closer to the results of
N-body simulations.  The fit of the NFW profile is characteristic using an
approximation given by del Popolo et al. (\cite{delpopolo}). However, in such an
approximation the final density profile depends only on the ``turnaround density
profile'' and not on the amount of angular momentum acquired by the system during
its expansion phase.


\section{The numerical method and the angular momentum}
The calculation of the collapse factor requires the evaluation of
the integral in (\ref{eqb6}). Changing the variables from the
turnaround radius to the initial one, this is written:
\begin{equation}
M_{add}(\zeta)=4\pi\rho_{b,i}\int_{x_i}^{x_b}P_{x_i}(x'_i)[1+\delta_i(x'_i)]x_i'^2\mathrm{d}x'_i,
\label{eqb10}
\end{equation}
where $P_{x_i}(x'_i)=I(x_i)/I(x'_i)$ with
\begin{equation}
I(r)=\int_{x'_p}^r\frac{1}{v_{g(x'_i)}(g(n))}\frac{\mathrm{d}g(n)}{\mathrm{d}n}\mathrm{d}n,\\
\label{eqb11}
\end{equation}
$\zeta = g(x_i)$, $x'_p=g^{-1}(x_p)$ where $x_p$ is the pericenter of the shell with
initial radius $x'_i$. The upper limit $x_b$ of the integral in (\ref{eqb10}), is
taken to be the initial radius of the sphere that has collapsed at the present
epoch. The radial velocity $v$ of a shell with apapsis $x=g(x_i)$ as it reaches the
radius $r=g(r_i)$ is given by the conservation of the energy of the shell and is:
\begin{equation}
v^2_x(r)=2[\Psi(r)-\varepsilon_x]-\frac{j^2_x}{r^2},\\
\end{equation}
where $\Psi$ equals minus the potential $\Phi$,$\varepsilon_x$
equals minus the specific energy and $j_x$ is the specific angular
momentum of the shell. The potential $\Psi$ after the change of
variables is given by the expression:
\begin{equation}
\Psi[g(r_i)]=\frac{GM(x_b)}{g(x_b)}+G\int_{r_i}^{x_b}\frac{M(x_i)}{g^2(x_i)}
\frac{\mathrm{d}g(x_i)}{\mathrm{d}x_i}\mathrm{d}x_i,\\
\end{equation}
where the distribution of mass $M(x_i)$ is that at the initial
conditions. The energy of the shell is calculated by:
\begin{equation}
\varepsilon_x=\Psi[g(x_i)]-\frac{j^2_x}{2g^2(x_i)}.
\end{equation}
The angular momentum is introduced by the following scheme:

Each shell expands radially from its initial radius $x_i$ up to
its maximum expansion radius $x$. At this stage a specific angular
momentum $j_x$ is added, given by
$j_x=\mathcal{L}\sqrt{M(x)x}=\mathcal{L}\sqrt{M(x_i)g(x_i)}$,
where $\mathcal{L}$ is a constant. This way of introducing angular
momentum is consistent with the angular momentum distribution in
N-body simulations (e.g, Barnes \& Efstathiou \cite{barnes}) and
does not introduce any additional physical scale. It has been used
by Avila-Reese et al. (\cite{avila}) and recently by Nusser
(\cite{nusser}).

In this  way the apocenter $r_a$ of a shell is its turnaround
radius $x$ while its pericenter $r_p$ is found by the solution of
the equation:
\begin{equation}
2r^2[\Psi(r)-\varepsilon_x]-j^2_x=0.
\end{equation}
The change of variables described above requires the solution for
r of the equation
\begin{equation}
2g^2(r)[\Psi(g(r))-\varepsilon_x]-j^2_x=0,
\end{equation}
where the potential $\Psi$ is that of the turnaround epoch.

Nusser (\cite{nusser}) proved the following two important
properties:
\begin{enumerate}
 \item If the angular momentum is introduced in the
above described way in a spherical system with a power law density
profile, then all shells have the same eccentricity. In fact, if
$\varphi \equiv r_p/r_a$ then the following equation holds;
\begin{equation}
G{\mathcal{L}}^2(\varphi^{-2}-1)=\frac{2[\Psi(r_p)-\Psi(r_a)]r_a}{M(r_a)}.
\end{equation}
The right hand side of the above equation can be expressed in
terms of $\varphi$ in the case of a power law density profile and
completes the proof.
\item
  If the potential evolves
adiabatically, given at time $t$ by the relation
$\Psi_t(r)=k(t)\Psi(r)$ with $k(t)$ a slowly varying function of
$t$ and $\Psi$ the potential at the turnaround epoch, and the
radial action is indeed invariant, then the eccentricity of every
shell remains constant during the evolution.

The radial action can be written in the form:
\begin{eqnarray}
J_r=j\times\nonumber\\
 \int_{\varphi}^1\left[(\varphi^{-2}-1)
\frac{\Psi(ur_a)-\Psi(r_a)}{\Psi(r_p)-\Psi(r_a)}
+(1-u^{-2})\right]^{\frac{1}{2}}\mathrm{d}u.
\end{eqnarray}
In the case of a power law density profile the quantity
$\frac{\Psi(ur_a)-\Psi(r_a)}{\Psi(r_p)-\Psi(r_a)}$ is written in
terms of $\varphi$ and $u$. Since $J_r$ is constant, then
$\varphi$ is constant.
\end{enumerate}
Nusser (\cite{nusser}) used these properties  to estimate the
asymptotic behavior of the density profile near and far from the
center of a system with a power-law initial density profile.
Unfortunately these properties do not hold for more realistic
density profiles. Similar power-law density profiles have been
used by Sikivie et al. (\cite{siki}) who used a CDM power spectrum
to estimate an ``effective'' exponent for this power-law on the
galactic scale. Sikivie et al. use a self-similar evolution of the
system in order to calculate its properties at the relaxed state.
Unfortunately, the self-similar evolution is not valid for more
realistic initial density profiles (non power-law profiles). The
numerical method presented in our study is more general. It is
able to deal with various initial density profiles (scale or not
scale free) and the assumption of self-similarity is not required.
Our results (presented in Sect.4) have derived for realistic
initial density profiles that have a finite value of the density
perturbation at the location of the peak. Therefore, these results
could be more reliable at least regarding the final state of the
central region of the system. Moreover, it is shown that taking
into account the angular momentum, final density profiles are well
fitted by two power law density profiles with slopes less than $2$
at the central regions of the systems and larger than $2$ at the
outer regions. This class of final density profiles is consistent
with the results of N-body simulations (eg Subramanian et al.
(\cite{subra})).

\begin{figure}[t]
\includegraphics[width=8.6cm]{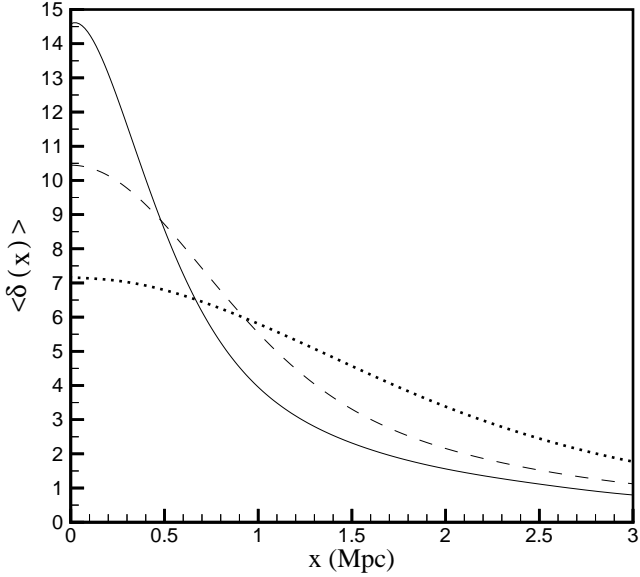}
\caption{Overdensity profiles versus the distance $x$ from a
$3\sigma$ peak. Dotted curve corresponds to a smoothing length of
$1.2 \mathrm{Mpc}$ while the dashed and solid curves correspond to
smoothing lengths of $0.6 $ and $0.3 \mathrm{Mpc}$ respectively.
The values of $<\delta (x)>$ and $x$ are both normalized to the
present.} \label{fig1}
\end{figure}
\section{The initial density profile and the results}
The averaged overdensity profile $<\delta(x)>$ at distance $x$
from a $n\sigma=n\xi^{1/2}(0)$ extremum of a smoothed density is
given in Bardeen et al. (\cite{bardeen}, hereafter BBKS) by the
equation:
\begin{eqnarray}
<\delta(x)> & =&\frac{n\xi(x)}{\xi(0)^{1/2}}\nonumber\\
-\frac{\theta(n\gamma,\gamma)}{\gamma(1-{\gamma}^2)}
[{\gamma}^2\xi(x)+\frac{R^2_*}{3}\nabla^2\xi(x)]/{\xi(0)}^{1/2},
\end{eqnarray}
where $\gamma\equiv I(4)/[I(2)I(6)]^{1/2}$ and
$R_*\equiv[3I(4)/I(6)]^{1/2}$ with
$I(l)=\int_0^{\infty}k^lP(k)\mathrm{d}k.$

In the above relations $\xi$ is the correlation function and $P$
the power spectrum. These are related by:
\begin{equation}
\xi(x)=\frac{1}{2{\pi}^2x}\int_0^\infty P(k)ksin(kx)\mathrm{d}k
\end{equation}

The function $\theta(n\gamma,\gamma)$ is given by the relation
\begin{equation}
\theta(n\gamma,\gamma)=\frac{3(1-{\gamma}^2)+(1.216-0.9{\gamma}^4)
e^{-\frac{\gamma}{2}(\frac {n\gamma}{2})^2}}
{[3(1-{\gamma}^2)+0.45+(\frac{n\gamma}{2})^{2}]^{\frac{1}{2}}+ \frac{n\gamma}{2}}
\end{equation}
in the range $1< n\gamma < 3$. The rms mass excess,$\frac{\Delta
M}{M}$, within a sphere of radius $x$, is given by:
\begin{equation}
\sigma_{x}=\frac{1}{(2\pi^{2})^{\frac{1}{2}}}\frac{3}{x^3}
\left[\int_{0}^{\infty}\frac{P(k)(\sin kx-kx\cos
kx)^{2}}{k^4}\mathrm{d}k\right]^{\frac{1}{2}}
\end{equation}
We used the spectrum calculated by BBKS for a CDM-dominated
Universe with $\Omega=1$ and $h=0.5$. This is given by the
equation
 \begin{equation}
 P_{CDM}=Ak^{-1}[\ln(1+4.164k)]^2[G(k)]^{-\frac{1}{2}}
 \end{equation}
 where $G(k)$ is the following polynomial of $k$.
 \begin{eqnarray}
 G(k)=192.9+1340k+1.599\times10^5k^2\nonumber\\
 +1.78\times10^5k^3+3.995\times10^6k^4
 \end{eqnarray}
 The above spectrum is smoothed on various scales according to the
 relation:
 \begin{equation}
 P(k)=P_{CDM}e^{-(\frac{k}{k_c})^2}
 \end{equation}
 Regarding the smoothing of the above spectrum three cases  are
 examined. In case A the spectrum is smoothed on a
  scale ${k_c}^{-1}=1.2$ Mpc, in case
 B on a scale ${k_c}^{-1}=0.6$ Mpc and in case C on
 ${k_c}^{-1}=0.3$ Mpc.
 The first scale length corresponds to a mass $0.5\times10^{12}M_{\sun}$ the
 second to a mass $6.25\times10^{10}M_{\sun}$ and the third to
 $7.8\times10^{9}M_{\sun}$.
 The constant of proportionality $A$ is chosen by the condition $\sigma_{8}=0.7$, (the rms
 mass excess within $8$ Mpc to be $0.7$). Then
 $<\delta(x)>$ is calculated for $n=3$ and plotted in
 Fig. \ref{fig1}. The dotted curve corresponds to the case A,
  the dashed curve to  the case B,
  while the solid one to the case C.
  We note that the values presented in this figure are
 normalized to the present.

 In the case of linear growth
 of the overdensity the following hold (Gunn \& Gott \cite{gunn72}):
 a shell with initial velocity equal to the Hubble flow and an initial
 comoving radius $x$ has expanded up to a maximum radius
  $r_{max}$ in a time $t_{ta}$  (given by (\ref{eqb3}),
  $\Omega_i=1$ in our case).
  This maximum radius is given by the relation
 \begin{equation}
r_{max}=x/\Delta, \label{eqb26}
\end{equation}
 while the collapse time of the shell,$t_c$, is related to the  age of the
 Universe,$t_0$, by
\begin{equation}
t_c=\frac{3\pi}{2}t_0{\Delta}^{-3/2}.\label{eqb27}
\end{equation}
 However calculating
\begin{equation}
\Delta=\frac{3}{{x}^3}\int_0^{x}<\delta(u)>u^2\mathrm{d}u,
\label{eqb28}
\end{equation}
and using (\ref{eqb27}) and (\ref{eqb28}) the collapse time of a
shell and its turnaround radius are found. In fact the condition
$t_c/t_0=1$ gives the value of $\Delta$ for the shell that
collapses today. Then $x$ is found by solving numerically
(\ref{eqb28}) and $r_{max}$ is calculated by (\ref{eqb26}). In
case A the mass inside the shell that collapses today is about
$9.8\times10^{12}M_{\sun}$ and the radius of maximum expansion is
$1130$ Kpc. For the case B the mass is $4.3\times10^{12}M_{\sun}$
and the radius of maximum expansion is $870$ Kpc while for the
case C the values are $2.1\times10^{12}M_{\sun}$ and $685$ Kpc
respectively. The values of $x$ resulting from the numerical
solution of (\ref{eqb28}) are $3.18$, $2.45$ and $1.93$ Mpc
respectively. Finally, the initial conditions at redshift $z_i$
can be derived by dividing both $x$ and $<\delta(x)>$ by $1+z_i$.
Using $1+z_i=1000$ the value of $x_b$ in (\ref{eqb10}) are $3.18,
2.45$ and $1.93$ Kpc for
 the cases A, B and C respectively.

The amount of the angular momentum in the system is adjusted by
the value of $\mathcal{L}$, (see Sect.3), and is measured by the
value of the dimensionless spin parameter
\begin{equation}
\lambda\equiv\frac{L|E|^{\frac{1}{2}}}{GM^{\frac{5}{2}}}
\end{equation}
where $L, E$ and $M$ are the total angular momentum, the total
energy and the total mass of the system respectively. The mean
value of $\lambda$ resulting from N-body simulations (e.g.
Efstathiou \& Jones \cite{efstathiou}, Barnes \& Efstathiou
\cite{barnes}) seems to be about 0.05. In our calculations we used
different values of $\lambda$ with a maximum of 0.12. The maximum
value of $\lambda$ corresponds to $\mathcal{L}=0.26$.

The resulting density profiles are fitted by a two-power law curve
of the form
\begin{equation}
\rho_{fit}(r)=\frac{\rho_c}{(\frac{r}{r_s})^{\beta}(1+\frac{r}{r_s})^{\mu}}
\end{equation}
where the fitting parameters $\rho_c$, $r_s$, $\beta$ and $\mu$
are calculated finding the minimum of the sum
\begin{equation}
S=\sum_{i=1}^{NP}\left(\frac{\log\rho_{SIM}(r_i)-\log\rho_{fit}(r_i)}{\log\rho_{SIM}(r_i)}\right)^2
\end{equation}
where $\rho_{SIM}$ are the predictions of SIM. NP is the number of
points where the density is found. We used 100 points equally
spaced on a log scale. The estimation of the above fitting
parameters is done using the unconstrained minimizing subroutine
ZXMWD of IMSL mathematical library. The quality of fit is very
good as can be seen in the following three Figs. \\
\begin{figure}[t]
\includegraphics[width=8.6cm]{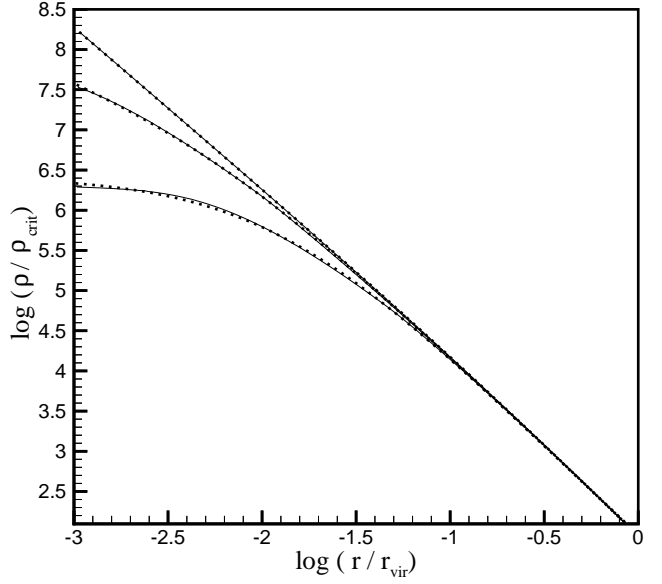}
\caption{Case A. Final density profiles derived for three
different values of the spin parameter $\lambda$. Solid curves:
SIM predictions. Dotted curves: the fits by a two-power law. From
the bottom of the figure the curves correspond to $\lambda = 0.09,
0.05$ and $0.0$  (radial collapse) respectively.} \label{fig2}
\end{figure}
\begin{figure}[]
\includegraphics[width=8.6cm]{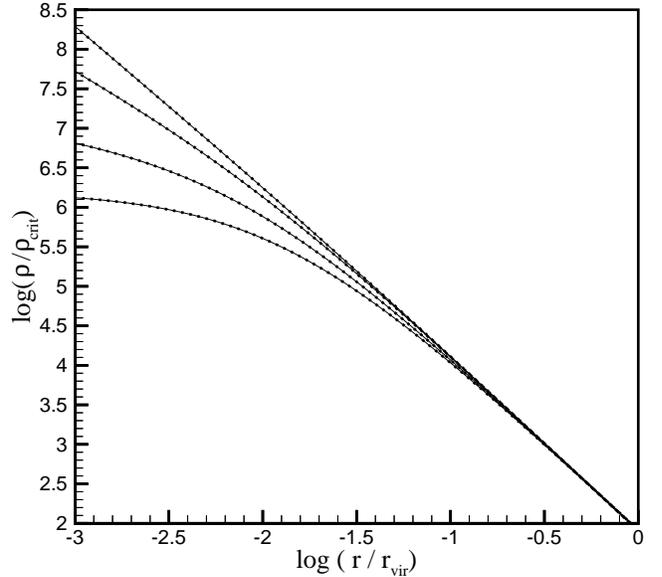}
\caption{Case B. As in Fig.\ref{fig2}. Solid curves: Density
profiles derived from the SIM. From the higher to the lower curve
the values of the spin parameter are $0.0, 0.05, 0.09 $ and $0.12$
respectively. Dotted curves: The fits of the solid curves by a
two-power law density profile.} \label{fig3}
\end{figure}
\begin{figure}[]
\includegraphics[width=8.6cm]{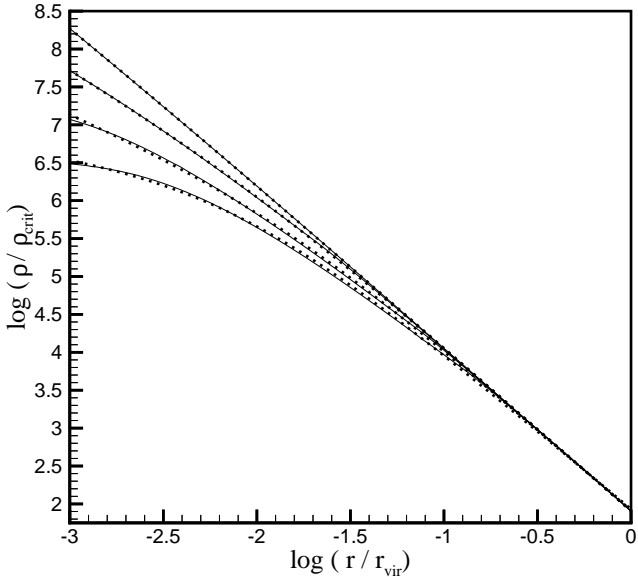}
\caption{ Case C. As in Fig.\ref{fig2}. From the higher to the
lower solid line the values of $\lambda$ are $0.0, 0.05, 0.09$ and
$0.12$ respectively. Dotted lines are the fits of the solid lines
by a two-power law density profile.} \label{fig4}
\end{figure}
In Fig. \ref{fig2} the final density profiles for the case A are
shown for three different values of $\lambda$. The radial collapse
($\lambda=0$) corresponds to the higher solid curve. The
intermediate curve corresponds to $\lambda=0.05$ while the lower
one to $\lambda=0.09$. Dotted lines are the fits by the above
described two-power law density profile. Distances are normalized
to the virial radius $r_{vir}$ which is $448$ kpc while densities
are normalized to the critical density $\rho_{crit}$. The virial
radius is the radius of the sphere with mean density $\approx 178$
times the present density of the Universe $\rho_{crit}$ (Cole \&
Lacey \cite{cole}). The virial mass of the system (the mass
contained inside the virial radius) is about
$4.7\times10^{12}M_{\sun}$. It is characteristic that the
efficiency of the angular momentum leads to shallower density
profiles in the inner regions of the system. Additionally at the
outer regions the density profile does not change even for the
maximum amount of angular momentum used. We note that for larger
values of the spin parameters the density profile becomes
unrealistic (increasing at the inner regions). As it will be shown
below this is a consequence of the shallow initial profile of this
case.

The results for the case B are presented in Fig. \ref{fig3}. The
values of the spin parameter are $0., 0.05, 0.09, $ and $0.12$.
The virial radius of the system is about $347$ $\mathrm{Kpcs}$ and
contains a mass of about $2.16\times10^{12}M_{\sun}$.

The case C is presented in Fig. \ref{fig4} for the same values of
$\lambda$ as in the case B. The virial radius of this system is
$273$ Kpc and its virial mass $1.06\times10^{12}M_{\sun}$.

Note that in the cases B and C the profile of the density
decreases even for larger values of $\lambda$ than used in  case
A, because of their steeper initial density profiles.

\begin{figure}[t]
\includegraphics[width=8.6cm]{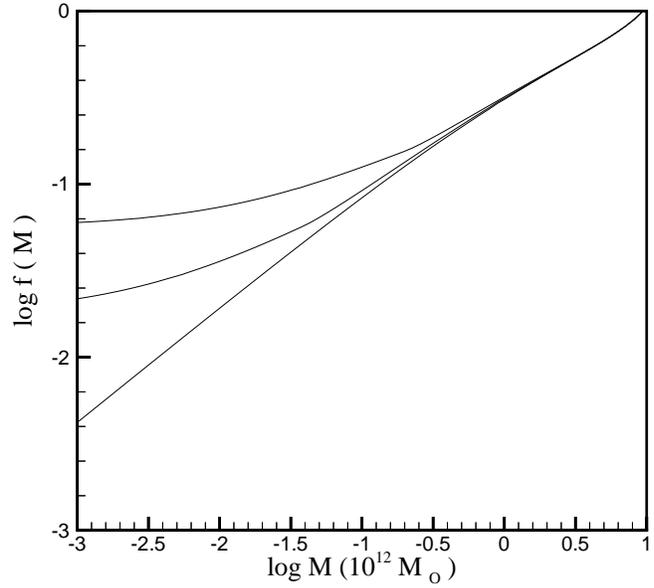}
\caption{ The collapse factor of mass $M$ versus $M$ on a
logarithmic scale for case A. From the lower line the values of
$\lambda$ are $0.0, 0.05, 0.09$ respectively.} \label{fig5}
\end{figure}

\begin{figure}[]
\includegraphics[width=8.6cm]{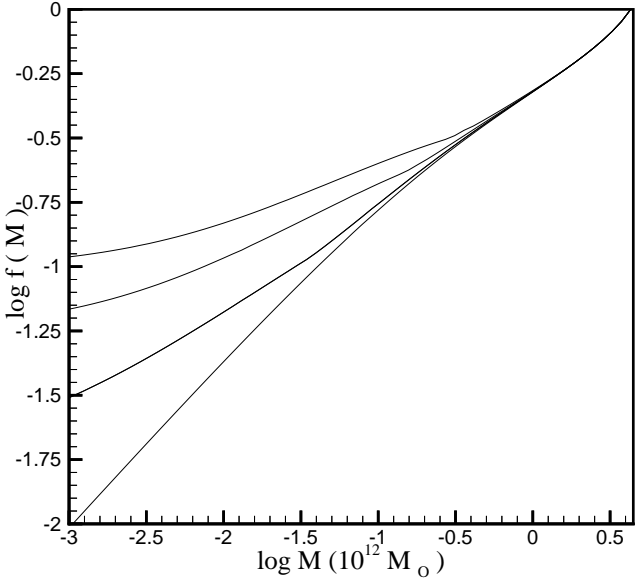}
\caption{ As in Fig.\ref{fig5} but for case B. The values of
$\lambda$, from the lower to the higher line are $0.0, 0.05, 0.09,
$ and $0.12$ respectively.} \label{fig6}
\end{figure}

\begin{figure}[]
\includegraphics[width=8.6cm]{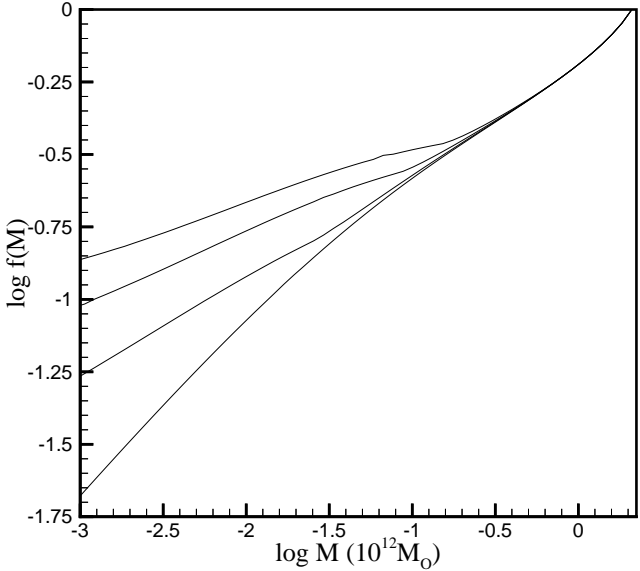}
\caption{ As in Fig.\ref{fig5} but for the case C.} \label{fig7}
\end{figure}

\begin{figure}[]
\includegraphics[width=8.6cm]{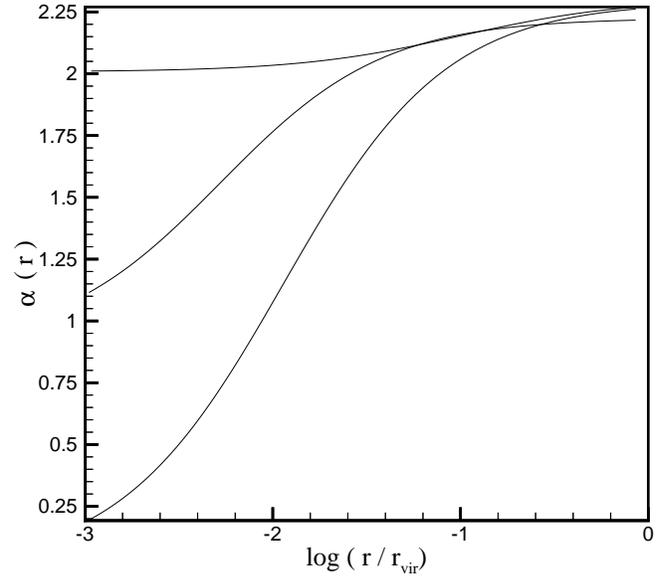}
\caption{ The slope of the final density profile versus radius for
case A. Distance is normalized to the virial radius. From the
lower to the higher line the values of $\lambda$ are $0.09, 0.05$
and $0.0$ respectively.} \label{fig8}
\end{figure}

\begin{figure}[]
\includegraphics[width=8.6cm]{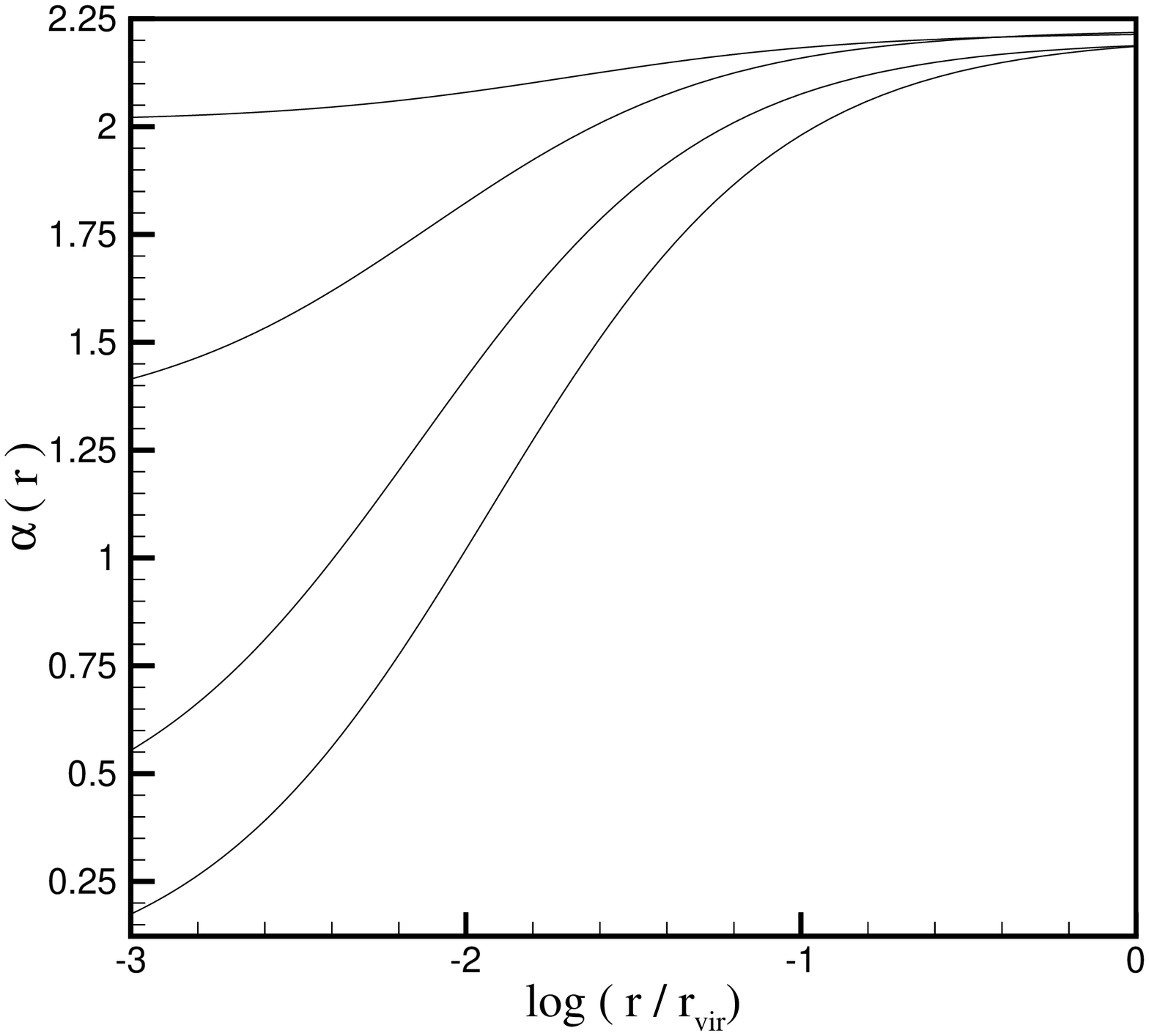}
\caption{ As in Fig.\ref{fig8} but for the case B. From the lower
to the higher line the values of $\lambda$ are $0.12, 0.09, 0.05$
and $0.0$ respectively.} \label{fig9}
\end{figure}

\begin{figure}[]
\includegraphics[width=8.6cm]{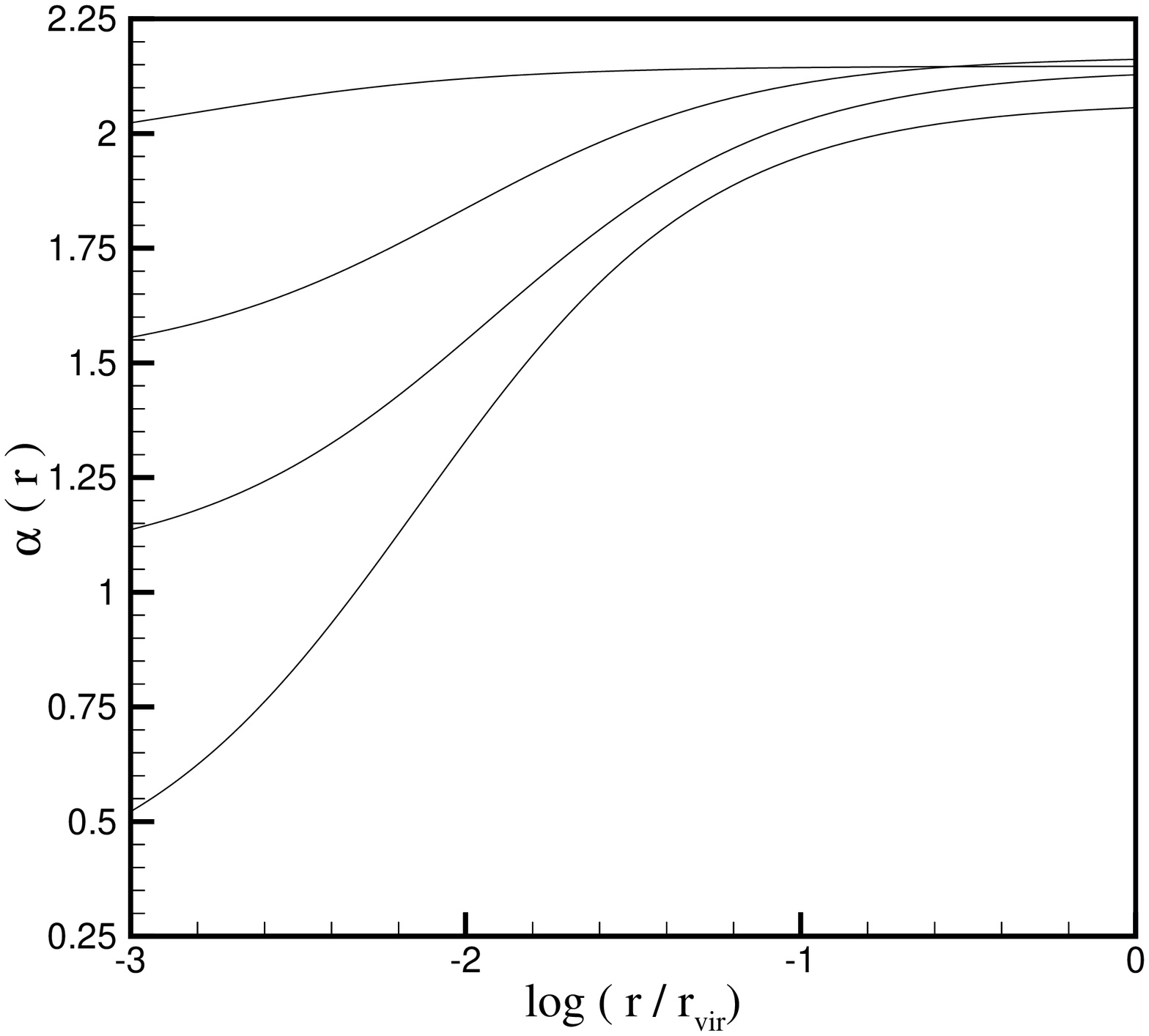}
\caption{ As in Fig.\ref{fig8} but for the case C. From the lower
to the higher line the values of $\lambda$ are $0.12, 0.09, 0.05$
and $0.0$ respectively.} \label{fig10}
\end{figure}

 In the following three figures the collapse factors for each case
 are presented. Fig. \ref{fig5} shows the collapse factor $f$ of
 mass $M$ on a logarithmic scale. The role of angular momentum is
 clear. Larger values of $\lambda$ lead to smaller values of $f$ and
 consequently to shallower density profiles as shown in Figs \ref{fig2},
 \ref{fig3} and \ref{fig4}. Fig.\ref{fig6} refers to case B and Fig.\ref{fig7}
 to case C. It is clearly shown in the above three figures that the collapse
 factor at the outer regions of the system is not affected by the amount
 of the angular momentum and it is almost the same  as that of the
 radial collapse case. The efficiency of angular momentum in creating shallow density
 profiles depends on the initial density profile. This can be shown in the
 following three Figs where the slope of the two-power law fit versus radius
 is plotted.  This is given by the relation
\begin{equation}
\alpha(r)=-\frac{\mathrm{d}\ln\rho_{fit}(r)}{\mathrm{d}\ln
r}=\beta+\mu\frac{\frac{r}{r_s}}{1+\frac{r}{r_s}}.
\end{equation}
Fig. \ref{fig8} corresponds to case A. The higher line is the
slope resulting after a radial collapse ($\lambda=0.0$) where the
values of $\alpha$ in the interval $0.001r_{vir}$ to $r_{vir}$ are
in the range $2.$ to $2.25$. The intermediate line corresponds to
$\lambda=0.05$ while the lower line corresponds to $\lambda=0.09$.
Fig.\ref{fig9} shows the slope for the case B. The lines, from the
higher to the lower, correspond to $\lambda=0.0, 0.05, 0.09$ and
$0.12$ respectively. The results of case C are shown in
Fig.\ref{fig10} for the same values of $\lambda$ as in case B. The
values of $\alpha$ for the three cases at $r=0.001r_{vir}$ are
clearly shown in the above figures. At $r=0.01r_{vir}$ and for
$\lambda=0.05$ the values of $\alpha$ are $1.76, 1.82$ and $1.83$
for the cases A, B and C respectively. A similar trend for the
inner regions --smaller $a$ for shallower initial density
profile--is also clear for all values of $\lambda$.  This trend is
reversed at the outer regions. The slopes at $r=r_{vir}$ are
approximately $2.25, 2.20$ and $2.15$ for the three cases
respectively. It is also clear from Fig.\ref{fig10} that the
radial collapse case leads to an almost exact power law profile.
In this case the slope at $r=0.01r_{vir}$ is $2.12$ while at
$r=r_{vir}$ is $2.15$.

\section{Conclusions}
The predictions of the SIM presented in this paper are summarized
as follows:

   \begin{enumerate}
      \item Radial collapse does not lead to power law final
      density profiles. However the difference in the slopes between the
      inner and the outer regions of the system is not large. Decreasing
      the smoothing scale of the power spectrum
      (leading to steeper initial profiles) this difference becomes
      smaller, leading to an almost power law for steep enough initial
      density profiles. The slopes in the radial collapse case are, in
      agreement with theoretical predictions (HS, Bertschinger
      \cite{bert}),
      in the range $2 $ to $2.25$.
      \item Angular momentum leads to shallower inner density
      profiles. The inner slope depends on the amount of the angular
      momentum, measured in our results by the value of the spin parameter,
      and on the form of the initial density profile. Angular
      momentum becomes more efficient, in decreasing $\alpha$, for shallower
      initial density profiles.
      \item The slope of density profiles does not change
      significantly at the outer regions of the system even in
      cases where a large amount of angular momentum is assigned
      to the system. At $r=r_{vir}$ the slope is approximately that of the radial
      collapse case.
\end{enumerate}
We note that the above results are limited by a large number of
assumptions, by the specific underlying cosmology and the
particular form of the power spectrum used. However, they show
systematic trends that could help us to better understand the
relation between the initial conditions and the final density
profiles. If things go the way described above, then the results
of N-body simulations could be approximated by adding angular
momentum to a case where the radial collapse results in a $r^{-3}$
density profile at the outer regions. However, the role of
different parameters of the problem is under study.

\begin{acknowledgements}
Thanks to the Empirikion Foundation for its support
\end{acknowledgements}

\end{document}